\documentstyle[epsfig,aps,multicol,prb]{revtex} 

\begin{document}
\author{Thorsten Dr\"ose and Cristiane  Morais-Smith}
\title{Metastability in Josephson transmission lines}

\address{ 
        I. Institut f\"ur Theoretische Physik, Universit\"at Hamburg,\\
        Jungiusstrasse~9, D-20355 Hamburg, Germany \\
        and \\
        Institut de Physique Th\'eorique, 
        Universit\'e de Fribourg,\\
        P\'erolles, CH-1700 Fribourg, Switzerland \\
        \rm{(\today)}
        }

\maketitle
\begin{abstract}
Thermal activation and macroscopic quantum tunneling in current-biased 
discrete Josephson transmission lines are studied theoretically. The degrees 
of freedom under consideration are the phases across the junctions which
are coupled to each other via the inductances of the system. The resistively 
shunted junctions that we investigate constitute a system of $N$ interacting 
degrees of freedom with an overdamped dynamics. 
We calculate the decay rate within exponential accuracy as a function of 
temperature and current. 
Slightly below the critical current, the decay from the metastable state
occurs via a unique ({\em rigid}) saddle-point solution of the Euclidean action
describing the simultaneous decay of the phases in all the junctions.
When the current is reduced, a crossover to a regime takes place, 
where the decay occurs via an {\em elastic} saddle-point
solution and the phases across the junctions leave the 
metastable state one after another.
This leads to an increased decay rate compared with the rigid case both
in the thermal and the quantum regime. 
The rigid-to-elastic crossover can be sharp or smooth analogous to first- or
second-order phase transitions, respectively.
The various regimes are summarized in a current-temperature decay diagram.
\end{abstract}

\begin{multicols}{2}
\section{Introduction}
The decay of systems in a metastable state 
has been intensively studied in the last decades. \cite{Haen90}
At high temperatures the decay is induced by thermal activation.
Lowering the temperature, under well-defined conditions quantum fluctuations 
become important and a crossover from thermal activation to 
quantum tunneling occurs. \cite{Affl81,Lark83,Grab84a,Weis99}
For systems with several interacting macroscopic degrees of freedom (DOF), 
the decay scenario is quite complex. For example, a crossover from
rigid decay, where all DOF decay instantly, to elastic decay,
where the DOF decay one after another \cite{Lefr92,Ivle87,Mora94} 
can take place in  systems with two DOF in the thermal \cite{Lefr92}
as well as in  the quantum decay regime. \cite{Ivle87,Mora94} 
Recently, 
the rigid-to-elastic crossover was investigated theoretically in 
systems with $N$ DOF  in the high-temperature limit. \cite{Droe99}
In this work we study the various decay regimes that occur in these
systems at low temperatures, where quantum tunneling is relevant.

A system in a  metastable state decays most probably via 
the lowest-lying saddle-point configuration of the Euclidean action. 
A crossover in the decay rate occurs, if a new lower-lying saddle point appears
upon tuning an external parameter. It can be continuous or discontinuous.
A continuous or second-order crossover occurs when 
a saddle-point bifurcation takes place.
For example, at the crossover from thermal to quantum decay, the 
static saddle point that determines the decay at high temperatures 
bifurcates into dynamic extrema (instantons) that have a lower Euclidean 
action. This causes an enhancement of the escape rate $ \Gamma $ for 
temperatures $T$ below the crossover temperature $T_{0}$ 
(see Refs.\ [2,3]).
In steepest-descent approximation $ \Gamma(T)$  and its derivative 
$ \Gamma'(T)$ are  continuous at $ T_{0}$, 
whereas the second derivative $ \Gamma''(T)$ diverges. 
In analogy to a classical phase transition, Larkin and Ovchinnikov 
denoted it a second-order crossover.\cite{Lark83} 
They pointed out that for some potential shapes $ \Gamma'(T)$ can become 
discontinuous at $T_{0}$ and the crossover is then of first order. 
Transitions of this type
were later investigated in more detail by Chudnovsky \cite{Chud92}
and studied in various physical systems by others. 
\cite{others,Goro97}
The crossover from rigid-to-elastic decay is similar and
both first-order and second-order crossovers have been found in systems
with two DOF. \cite{Ivle87,Mora94}

The DOF we study in this paper are macroscopic and coupled to their 
environment. At low temperatures, when macroscopic quantum tunneling 
(MQT) occurs, the interaction of the DOF with the environment
leads to quantum dissipation. \cite{Cald81}  Among the experiments 
that have been successfully 
interpreted in the framework of MQT with dissipation 
are the investigations of the decay of metastable states in current-biased 
Josephson junctions (JJ's)  and superconducting quantum 
interference devices (SQUID's). \cite{Voss81}
The relevant collective coordinate
in the case of the JJ is the phase difference across the junction.
The dc SQUID consists of two parallelly connected JJ's and thus 
enables the study of the quantum dynamics of two coupled degrees of freedom,
the phases across the junctions. The natural generalization to $ N $
degrees of freedom are parallel coupled one-dimensional 
Josephson-junction arrays, also known as discrete Josephson transmission 
lines (DJTL's). The continuum limit $ N \to \infty $ 
corresponds to a long JJ if the length of the junction $ l $ is larger than 
the Josephson length $ \Lambda_{J}$. The thermal decay of the phase in 
underdamped long JJ's (Refs. \cite{Sima90,Cast96,Sima97}) and 
overdamped DJTL's 
(Ref. \cite{Droe99}) has been investigated recently. The quantum tunneling in 
underdamped long JJ's has been analyzed in the limit $ l \to \infty $  
(Ref. [15]) and in the rigid regime $ l {\mathrel{\raise.4ex\hbox{$<$}
\kern-0.8em\lower.7ex\hbox{$\sim$}}} l_{c}$, where 
$ l_{c} \sim \Lambda_{J} $ is the critical length above which elastic decay 
via boundary nucleation sets in. \cite{Sima97} To date, the interesting case 
$ l {\mathrel{\raise.4ex\hbox{$>$}\kern-0.8em\lower.7ex\hbox{$\sim$}}} l_{c}$ 
has not been treated in the quantum-tunneling regime,
neither in the overdamped nor in the underdamped case. Solving the problem 
is difficult since then the saddle-point solutions of the action are 
inhomogeneous in time {\em and} space.   
Two reasons motivated us to study $ N $ harmonically coupled DOF trapped in a 
metastable state. First, our model can be used to describe the decay of the 
JJ phases in a DJTL and in the 
continuum limit $ N \to \infty$ to represent a long JJ. 
The second reason is the need to analyze the 
so far undiscussed overdamped limit. 

Since the decay close to the rigid-to-elastic crossover is determined
by the long-wavelength excitations of the system, much insight can be 
gained from studying dc SQUID's. Experiments on dc SQUID's 
(Refs. \cite{Yong86,Shar88,Han89}) showed that at high currents
the classical decay mainly proceeds via the simultaneous activation
of both phases across a common saddle point in the potential-energy 
landscape. However, below a crossover current, it was found that 
this saddle point splits and the phases decay one after another. \cite{Lefr92}  
A similar phenomenon was proposed to occur in the quantum decay process
in underdamped \cite{Ivle87} and overdamped \cite{Mora94} SQUID's,
known as the instanton splitting. In this case, the common saddle point 
of the Euclidean action bifurcates into two saddles of lower action.
In addition, a regime was found, where upon tuning temperature or current
a first-order transition takes place before the saddle-point bifurcation
occurs.  

The rigid-to-elastic crossover in DJTL's is similar. Of course, instead of two 
DOF, one then has $N$ coupled DOF and the theory has to be generalized 
accordingly. A perturbative treatment to calculate the elastic instanton 
solutions of the Euclidean action has been sketched in Refs.\ [7,8] 
for case of the dc SQUID. In the present publication, the 
perturbation scheme is systematically worked out  and applied to the problem 
of an overdamped DJTL in a metastable state. With this procedure
one is able to  calculate the split-instanton solution
to arbitrary precision. We  present 
{\em quantitative} results for the quantum decay rate of overdamped dc SQUID's,
DJTL's and long JJ's close to the the rigid-to-elastic crossover. 
We further construct the decay diagram of DJTL's and long JJ's  
and compare it to the one of dc SQUID's.

The paper is organized as follows: In Sec.\ II we introduce
the Euclidean action to describe the quantum decay in DJTL's with dissipation. 
In Sec.\ III, we discuss the saddle-point solutions of the action as a function
of the temperature and the driving current. An iterative 
perturbation procedure to calculate these extrema close to a saddle-point
bifurcation is presented in Sec.\ IV.  This method is then 
applied to evaluate the split-instanton solutions in Sec.\ V. 
The various decay regimes, which are summarized in a diagram 
and the corresponding relaxation rates are discussed in Sec.\ VI. 
Finally, the conclusions are  drawn in Sec.\ VII.

\section{Model}
\subsection{Decay rate}
The probability per unit time for a 
metastable state to decay at a finite temperature $ T $ is related to the 
imaginary part of the free energy $ F$ by 
$ \Gamma = (2/\hbar) {\rm Im} F$, as was shown by Affleck. \cite{Affl81}
In the path-integral representation, the free energy 
is given by
 $ F = - kT \ln [\oint {\cal D}[{\bf q}] \exp(-S_{E}[{\bf q}]/\hbar) ] $,
where 
$ S_{E}$ is the Euclidean action and $ {\bf q}$ is a vector representing
the DOF. If the energy barrier that the system has to overcome in order to 
leave the metastable state is
large, the sum over paths is dominated by the extremal trajectories 
and the latter can be treated in the  semiclassical approximation.
The decay rate  then reads
\begin{equation}
\Gamma = A e^{-B/\hbar},
\end{equation}
where $ B $ is the Euclidean action $ S_{E}$ evaluated at the extremal 
saddle-point trajectory and $ A $ is  a prefactor that can be obtained 
by considering the fluctuations around the extremal path. In this work,
we will determine the exponent $ B$.

\subsection{Euclidean action}
The Euclidean action of a system of $ N $ identical JJ's in the 
presence of a bias current $ I $ is
\begin{displaymath}
S_{E} =  
\int_{0}^{\hbar/T} d {t} \left[
K( \mbox {\boldmath$\varphi$})+ V( \mbox {\boldmath$\varphi$})+ D(\mbox {\boldmath$\varphi$})  \right],
\end{displaymath}
where $ {t} $ is the imaginary time,
$ \mbox {\boldmath$\varphi$} = (\varphi_{0},\dots,\varphi_{N-1})$ 
represents the phase differences across the junctions, 
$V$ is the potential energy, and $D$ models the dissipation.
The ``kinetic'' energy containing the  capacitances  
$ C$ of the JJ's is given by 
\begin{displaymath}
K( \mbox {\boldmath$\varphi$}) =  
 \frac{m}{2} \sum_{n=0}^{N-1} 
\left( \frac{\partial \varphi_{n}}{\partial {t} } \right)^{2},
 \end{displaymath}
where $m=C(\Phi_{0}/2\pi)^{2}$ corresponds to the ``mass'' of a fictitious 
particle and $\Phi_{0}=hc/2e $ is the 
flux quantum.
The potential energy consists of two parts, 
\begin{displaymath}
V(\mbox {\boldmath$\varphi$}) = U(\mbox {\boldmath$\varphi$}) + E(\mbox {\boldmath$\varphi$}), 
\end{displaymath}
where 
$  U( \mbox {\boldmath$\varphi$}) = 
E_{J} \sum [1-\cos(\varphi_{n})-(I/N I_{c}) \varphi_{n}]  
$
represents the tilted washboard potential of the driven JJ's  
that arises due to the relation between currents and gauge invariant phases  
across the junctions. Here $I$ is the
total current through the system, $ I_{c} $ is the critical current of a 
single junction, and  $ E_{J}=(\Phi_{0}/2\pi)I_{c}$ 
is the Josephson energy. 
We concentrate on the experimentally most 
interesting limit of currents close to criticality, 
$ N I_{c} - I \ll I_{c} $,
where the tilted washboard potential can be well approximated by its 
cubic expansion,
\begin{displaymath}
U(\mbox {\boldmath$\varphi$}) = U_{0} + \frac{E_{J}}{2} \sum_{n=0}^{N-1} 
 \left[ \epsilon \left(\varphi_{n}- \tilde{\varphi} \right)^{2}
  - \frac{1}{3}\left(\varphi_{n}- \tilde{\varphi}  \right)^{3} \right]. 
\end{displaymath}
Here  $\epsilon = \sqrt{2(1-I/NI_{c})}$ is a small parameter that
indicates the distance from criticality $I=NI_{c}$, 
$ \tilde{\varphi} = \pi/2 -\epsilon $  is the value of the phase at the 
minimum of the potential, 
and $ U_{0} $ is an irrelevant constant that
will be dropped in the following. 
Taking only the self-inductances $ L$ of the loops into account and 
neglecting the mutual inductances \cite{Bock94}, 
the interaction energy between the 
loops is 
\begin{displaymath} 
E(\mbox {\boldmath$\varphi$}) =  \frac{E_{J}}{2 \beta} 
\sum_{n=0}^{N-2} (\varphi_{n+1} -\varphi_{n})^{2},
\end{displaymath}
where $ \beta = L I_{c}^{2}/ E_{J} $ is the McCumber parameter.
We model dissipation due to an Ohmic 
shunt resistance $R$ using the Caldeira-Leggett approach \cite{Cald81},
\begin{displaymath}
D(\mbox {\boldmath$\varphi$}) = -\frac{\eta}{2 \pi} 
\int_{0}^{\hbar/T} d {t}'  
\frac{\partial \mbox {\boldmath$\varphi$} }{\partial {t}' }
\cdot \frac{\partial  \mbox {\boldmath$\varphi$}}{\partial {t} }
\ln \left| \sin \left[ \frac{\pi T}{ \hbar }({t}-{t}') \right] \right|,
\end{displaymath}
where $ \eta=1/R $ is the phenomenological friction coefficient.
In the following we will treat the overdamped limit
and neglect the contribution of the capacitive term 
$ K(\mbox {\boldmath$\varphi$}) $ to the action.
It is convenient to perform a transformation to 
dimensionless normal coordinates,
\begin{eqnarray} \label{trans}
\varphi_{n}({t})& = & {\tilde \varphi} + 2 \epsilon q_{0}(\tau) \nonumber \\
&& + 2 \sqrt{2}~ \epsilon  \sum_{k=1}^{N-1} q_{k}(\tau) 
\cos \left( \frac{\pi k \left( n+ 1/2 \right)}{N} \right). 
\end{eqnarray}
Note that the $ q_{k} $ are functions of the 
dimensionless imaginary time $ \tau = 2 \pi T {t} / \hbar - \pi $.
We now define the dimensionless potential energy, 
\begin{equation} \label{dimlesspot}
{\cal V}({\bf q}) = \frac{1}{2 N E_{J} \epsilon^{3}} 
V( \mbox {\boldmath$\varphi$}) 
=    
\frac{1}{2}\sum_{k=0}^{N-1} \mu_{k} q_{k}^{2}  +  {\cal N}({\bf q}), 
\end{equation}
where
 \begin{equation} \label{eigen_min}
\mu_{k} =
 \frac{8}{\beta \epsilon} 
\sin^{2} \left( \frac{\pi k}{2 N} \right) + 2 
\end{equation}
are the eigenvalues of $(\partial_{m} \partial_{n}{\cal V}) $
evaluated at the local  minimum $ {\bf q} = {\bf 0}$.
${\cal N}({\bf q})  $ contains the cubic terms 
\begin{eqnarray}
{\cal N}({\bf q}) 
& = & - \frac{2}{3} q_{0}^{3}  - 2 q_{0} \sum_{k=1}^{N-1} q_{k}^{2}   
-\frac{\sqrt{2}}{3}  
\sum_{k=1}^{N-1}q_{k}^{2}  \left( q_{2k} - q_{2(N-k)} \right) 
\nonumber \\
& &  -\frac{2 \sqrt{2}}{3}   \sum_{m>k=1}^{N-1} 
     q_{m} q_{k} \left( q_{m+k} +q_{m-k} - q_{2N-m-k}\right)  \nonumber
\end{eqnarray}
with $ q_{k}=0$ for $k>N-1$.

The action in terms of the dimensionless coordinates reads
\begin{equation} \label{action_q}
S_{E} = g \int_{-\pi}^{\pi} d \tau L_{E},
\end{equation}
where the Euclidean Lagrangian is 
\begin{displaymath}
L_{E}  =  {\cal V}( {\bf q})
 - \frac{\theta}{\pi {\sqrt {J}}} \frac{\partial {\bf q}}{\partial \tau}
   \int_{-\pi}^{\pi} d \tau'  \frac{\partial {\bf q}}{\partial \tau'}
\ln \left| \sin \left[ \frac{\tau-\tau'}{2} \right] \right|.
\end{displaymath}
Here $ g = N  \hbar E_{J} \epsilon^{3} / \pi T \gg 1$
is the semiclassical parameter,
$  \theta = \pi \eta \beta T / 2 \hbar E_{J} \sin^{2}(\pi/2N) $ 
is the dimensionless temperature, 
and $  {J} = (\beta \epsilon)^{2}/[2 \sin(\pi/2N)]^{4}$ 
is the dimensionless 
current. Note that $ {J} =0$ corresponds to $ I=NI_{c}$.

\section{Extrema of the Euclidean action}
By applying the variational principle to the Euclidean action, one finds the
classical equations of motion in imaginary time
\begin{equation} 
\nabla_{q} {\cal V}({\bf q})
+  \frac{ \theta}{\pi {\sqrt {J}}}
   \int^{\pi}_{-\pi} d \tau'  \frac{\partial {\bf q} }{\partial \tau'}
\cot \left[ \frac{\tau-\tau'}{2}   \right] = 0.
\end{equation}
Their solutions are given by the extremal trajectories, of which the 
saddle-point solutions are of special interest, since they lead to the
decay of the chain from its metastable state. 
At high temperatures, quantum fluctuations play a minor role and the
solutions of the equation of motion are time independent,
$ \partial_{\tau}{\bf q}=0 $. Hence, they
are given by the extrema of the potential, $\nabla_{q} {\cal V}({\bf q})=0 $. 
However, below a crossover temperature $ \theta_{0}$,
quantum tunneling becomes relevant for the decay process and the solutions
of the equation of motion are a function of the imaginary time. 
In the following paragraphs, we first analyze the extrema of the potential,
which determine the decay in the thermal regime.
Then we derive an expression for the crossover 
temperature $ \theta_{0} $ from 
thermal to quantum tunneling and finally analyze the time-dependent extremal
solutions of the Euclidean action, the instantons which lead to decay
via quantum tunneling.

\subsection{Saddle points of the potential energy}
A trivial type of saddle-point solution can be readily constructed from 
physical arguments. Consider the case where the attractive interaction 
between the particles is much larger than the energy barrier. 
At high temperatures,  the strongly coupled particles 
most probably are thermally activated over the barrier all at once and
the chain basically behaves as a {\em rigid} rod. 
In this case the saddle point of the potential  $ V $ is identical to the 
local maximum of the single DOF potential, $ \varphi_{0} = \dots = 
\varphi_{N-1} 
= \epsilon + \pi/2$, which reads $ q_{0} = 1, q_{k>0} = 0$ in normal mode 
representation. If, on the other hand, the energy barrier
is of the order of the interaction strength or even larger, 
another saddle point of the potential emerges. Then 
above a certain barrier height the chain preferably decays via a kinked 
saddle-point solution with $ q_{k>0} \not= 0 $ that we call {\em elastic}.

Taking the second derivatives of Eq.\ (\ref{dimlesspot}) 
at $ {\bf q}_{rs} = (1,0,\dots,0)$ we evaluate
the eigenvalues of the curvature matrix at the thermal rigid saddle point,
\begin{equation} \label{eigen_sad}
\lambda_{k}^{rs} = \partial_{k}^{2} {\cal V}( {\bf q}_{rs})
= \frac{8}{\beta \epsilon} 
\sin^{2} \left( \frac{\pi k}{2 N} \right) - 2. 
\end{equation}
Note that $ \lambda_{1}^{rs}= 2(1/\sqrt{J}-1)$
becomes negative for $ J > 1$, indicating that another saddle appears.
This {\em elastic} solution has a lower 
activation energy and hence becomes the most probable configuration
leading to escape from the metastable state. \cite{Droe99}

To determine the saddle points,
one has to solve a system of $ N $ coupled nonlinear equations.
So far, elastic saddles have been calculated exactly for $ N=2$, 
Ref.\ [6] and $ N=3$, Ref.\ [9]. 
Close to the crossover  
$ J {\mathrel{\raise.4ex\hbox{$>$}\kern-0.8em\lower.7ex\hbox{$\sim$}}} 1 $, 
the $ q_{k>0}$ are small. 
Hence, for $ N>3$,  approximate solutions can be found. \cite{Droe99}
Expanding around the rigid saddle, $ q_{0} = 1 + {\tilde q}_{0} $ and
$ q_{k} = {\tilde q}_{k}$ for $ k>0$, we approximate the potential energy by
\begin{equation} \label{Vapprox}
{\cal V} =
\frac{1}{3}
+ \frac{1}{2} \sum_{k=0}^{N-1} \lambda_{k}^{rs} {\tilde q}_{k}^{2}  
- {\tilde q}_{1}^{2} \left( 2 {\tilde q}_{0} + \sqrt{2} {\tilde q}_{2} 
\right). 
\end{equation}
Solving $ \nabla_{q} {\cal V}=0$, one finds 
${\tilde q} _{1}^{2} = 
\lambda_{0}^{rs} \lambda_{1}^{rs} \lambda_{2}^{rs}/(4 \lambda_{0}^{rs} 
+ 8 \lambda_{2}^{rs} )$,
$ {\tilde q}_{0} = 2 {\tilde q}_{1}^{2}/\lambda_{0}^{rs}$,
$ {\tilde q}_{2} = \sqrt{2} {\tilde q}_{1}^{2}/\lambda_{2}^{rs}$,
and $ {\tilde q}_{k>2} = 0$.
Hence the eigenvalues of the curvature matrix 
evaluated at the elastic saddle  for $ k \not= 1$ are 
$ \lambda_{k}^{es} \sim \lambda_{k}^{rs} $, whereas 
$\lambda_{1}^{es} = 2 | \lambda_{1}^{rs}|.$
Inserting these results into Eq.\ (\ref{Vapprox}), one finds that the 
activation energy is reduced due to the elasticity. Close to the crossover,
where 
$ J {\mathrel{\raise.4ex\hbox{$>$}\kern-0.8em\lower.7ex\hbox{$\sim$}}} 1$, it 
is given by
\begin{equation} \label{VCO}
{\cal V} \approx
\frac{1}{3} - \frac{|\lambda_{0}^{rs}| \lambda_{2}^{rs}}
                   {4(\lambda_{0}^{rs}+2\lambda_{2}^{rs})} 
\left(\frac{1}{\sqrt{J}} -1 \right)^{2}.
\end{equation}

\subsection{Crossover temperature from thermal to quantum decay}
In the following, we derive the crossover temperature $ \theta_{0}$.
Since the time-dependent quantum saddle-point solutions  
are periodic in $ \tau $ with a period $ 2 \pi $, 
they can be represented by a Fourier series. Near the crossover, 
the Fourier series can be well approximated by
$ {\bf q}(\tau) \approx {\bf q}_{ts} + {\bf p} \cos(\tau)$,
where $ {\bf q}_{ts} $ is the thermal saddle point under consideration
(rigid or elastic) and
$ {\bf p} $ is a small correction term due to quantum fluctuations. 
Substituting  $ {\bf q}(\tau) $ into the linearized equation of motion, 
\begin{displaymath}
 \left[ \partial_{k} \partial_{l} {\cal V}({\bf q}_{ts}) \right]
 {\bf p} \cos(\tau)  =   \frac{\theta {\bf p} }{\pi {\sqrt {J}}}
   \int d \tau' \sin(\tau') 
\cot \left( \frac{\tau-\tau'}{2} \right) 
\end{displaymath}
one finds an eigenvalue equation for the potential curvature matrix,
\begin{displaymath}
\left[ \partial_{k} \partial_{l}  {\cal V}({\bf q}_{ts}) \right]
{\bf p} 
= - \frac{2 \theta}{{\sqrt {J}}}  {\bf p}.
\end{displaymath}
The only negative eigenvalue of the curvature matrix 
$\left[ \partial_{k} \partial_{l} {\cal V}({\bf q}_{ts}) \right] $
evaluated at the saddle point of the potential is given by $ \lambda_{0}$,
the curvature along the unstable direction.
Hence the crossover temperature is given by
\begin{equation} \label{T_0}
\theta_{0} =  \frac{ |\lambda_{0}| {\sqrt {J}}}{2}.
\end{equation}
For the rigid regime, where $ \lambda_{0}^{rs}=-2$ we thus find
\begin{equation} \label{T_0:rigid}
\theta_{0}(J< 1) =  {\sqrt {J}},
\end{equation}
and in the elastic regime with 
$ \lambda_{0}^{es} \approx  \lambda_{0}^{rs} - 
2 \lambda_{1}^{rs} \lambda_{2}^{rs}/(\lambda_{0}^{rs} -\lambda_{2}^{rs} )   $
\begin{equation} \label{T_0:elastic}
\theta_{0}(J 
{\mathrel{\raise.4ex\hbox{$>$}\kern-0.8em\lower.7ex\hbox{$\sim$}}} 1)  = 
\sqrt{J} + \left( J-\sqrt{J} \right) 
\left(2-\frac{1}{2 \cos^{2}(\pi/2N )} \right).
\end{equation}

\subsection{Instantons}
For  $ \theta < \theta_{0} $ quantum tunneling becomes relevant  and the
instanton solutions dominate the decay from metastability.
For $ J>1$, the crossover from thermal to quantum 
decay is of second order. A detailed procedure showing how to obtain the 
saddle-point solutions close to the thermal-to-quantum crossover was given in 
Ref.\ [12] and can be applied to calculate the quantum elastic solutions.
In this paper we will concentrate on the current regime  $ J<1$,
where in addition to the transition from the thermal to the quantum rigid
regime a crossover from the quantum rigid to the quantum  elastic phase can 
take place. 

The rigid quantum solution is found by setting
$ q_{k} = 0 $ for $ k>0$. In this case the equations 
$ \delta S_{E} / \delta q_{k} = 0 $ with $ k>0$
are trivially satisfied and the remaining equation for $ k=0 $ describes
the thermally assisted quantum tunneling of a single degree of freedom 
$ q_{0}$. Its solution is the well-known instanton obtained by Larkin and 
Ovchinnikov \cite{Lark83}
\begin{equation}
q_{0}^{(0)}(\tau) =  \left( \frac{\theta}{\theta_{0}} \right)^{2} 
\frac{1}{1-{\sqrt{1-\theta^{2}/\theta_{0}^{2}}~ \cos(\tau) }}.    
\end{equation}
Inserting $ q_{0}^{(0)}$ and $q_{k>0}^{(0)}=0 $ into Eq. (\ref{action_q}),
one obtains the extremal action in the rigid quantum regime
\begin{equation} \label{Bqr}
B_{qr} = B_{0} 
\left[1- \frac{1}{3}\left( \frac{\theta}{\theta_{0}} \right) ^{2} \right],
\end{equation}
where $ B_{0} = 2 \pi N \eta \epsilon^{2} $.

As for  $J>1$, nonuniform saddle-point solutions of 
the action exist in the quantum regime in a certain parameter range
even for $J<1$.
If the action evaluated at this extremum is lower than $ B_{qr}$,
the nonuniform configuration is the most probable one leading
to  decay from the metastable state via quantum tunneling. 
Tuning the temperature $ \theta $ at a fixed bias current, a 
nonuniform saddle-point solution can develop in two different ways.
One possibility is that  a less probable nonuniform configuration,
which coexists with the rigid saddle point above a critical temperature 
$ \theta_{1}$, 
becomes the lowest-lying saddle point of the Euclidean action below 
$ \theta_{1} $. Then the most probable configuration abruptly changes
from uniform to nonuniform. Since the first derivative of the rate 
$ \partial_{\theta} \Gamma $ is discontinuous at $ \theta_{1}$, the crossover
is of first order. Another scenario is encountered, if at a critical
temperature $ \theta_{2} $ the rigid saddle point bifurcates into new
saddle points which have the lowest action. 
This crossover, known as instanton splitting,
\cite{Ivle87,Mora94} is of second order. 

A strategy to determine nonuniform saddle-point solutions for $J<1$
is to first search for a saddle-point splitting and then to verify 
whether a first-order transition might have occurred before the
bifurcation has taken place. 
If a first-order transition can be ruled 
out, the new bifurcated saddle points have the lowest action. In this 
case the bifurcation causes a second-order crossover from a single 
to a split-instanton regime.

Following this idea, we first identify the saddle-point bifurcation,
calculate the split instantons and test whether or not a first-order
transition has already occurred.

\section{Iterative perturbation scheme}
In this  section we present an iteration scheme to 
calculate the split-instanton solutions for $J<1$.
We start by 
expanding  the coordinates around the single instanton solution,
\begin{equation}
q_{k}(\tau) = q_{k}^{(0)}(\tau) + {\tilde q}_{k}(\tau).
\end{equation}
and rewrite the Euclidean action in terms of the new variables,
\begin{eqnarray} \label{action2}
S_{E} & = & B_{qr} + g  \int_{-\pi}^{\pi} d\tau 
\left[ \frac{1}{2}\sum_{k=0}^{N-1} 
{\tilde q_{k}} {\hat Q}_{k} {\tilde q_{k}} + {\cal N}({\tilde {\bf q}})
\right].
\end{eqnarray}
The operators $ {\hat Q}_{k} $ are defined as
\begin{eqnarray}
{\hat Q}_{k} {\tilde q}_{k}(\tau)  &=& \frac{1}{g}  
\int_{-\pi}^{\pi} d\tau' 
\frac{\delta^{2} S_{E}[{\bf q}^{(0)}]}
{ \delta q_{k}(\tau){\delta q_{k}(\tau') }} 
{\tilde q}_{k}(\tau') \nonumber \\
&=& \
(\mu_{k} -4 q_{0}^{(0)})
{\tilde q}_{k}(\tau) \nonumber \\
&&+  \frac{\theta}{\pi {\sqrt {J}}}
   \int_{-\pi}^{\pi} d \tau'  \frac{\partial {\tilde q}_{k}(\tau')}
                                   {\partial \tau'}
\cot \left( \frac{\tau-\tau'}{2} \right) \nonumber.
\end{eqnarray}
To determine the split instantons, we have to find saddle points of
the action with nonzero ${\tilde q}_{k}$. 
Hence we have to solve the equations of motion, 
\begin{equation} \label{pertequ}
{\hat Q}_{k}{\tilde q}_{k}   = 
- \frac{\partial{\cal N}({\tilde {\bf q}})}{\partial{\tilde q}_{k}},
\end{equation}
which constitute a system of coupled nonlinear differential 
equations. In general, they cannot be solved exactly. 
However, close to the saddle-point bifurcation, the extremal amplitudes
$ {\tilde q}_{k} $ are small and
we can calculate
approximate solutions by applying an iterative perturbation scheme.
This leads to a  hierarchy of inhomogeneous {\em linear} equations,
\begin{equation} \label{inhomo} 
{\hat Q}_{k}{\tilde q}_{k}^{(i)}   = F_{k}^{(i)}, 
\end{equation}
where $i$ denotes the iteration step.
In the first iteration $i=1$ 
we take only terms into account that are linear in 
${\tilde q}_{k}$. The  higher-order terms on the right-hand side of 
Eq.\ (\ref{pertequ}) are neglected. Thus,  $ F_{k}^{(1)} =0$ 
and Eqs.\ ({\ref{inhomo}}), which have to be solved, are homogeneous.
For $ i>1 $, the amplitudes calculated in the previous iteration are
substituted into $ \partial {\cal N} $ such that  
the inhomogeneous terms are given by
\begin{equation} \label{force}
 F_{k}^{(i)}(\tau) = 
-\frac{ \partial {\cal N}[{\tilde {\bf q}}^{(i-1)}(\tau)]}
      { \partial {\tilde q}_{k} }.
\end{equation}
After each iteration step $ i $, we thus obtain approximate 
(special) solutions for the amplitudes ${\tilde q}_{k} $
by formally inverting Eq.\ (\ref{inhomo}),
\begin{equation}  \label{inverseinhomo}
{\tilde q}_{k}^{(i)}   ={\hat Q}_{k}^{-1} F_{k}^{(i)}. 
\end{equation}
Of course, a straightforward inversion is not possible, if 
$ {\hat Q}_{k} $ is singular. Below, we will discuss how to 
handle equations with a singular operator.

The inversion is most conveniently 
performed by representing Eq.\ (\ref{inverseinhomo}) in terms 
of the  eigenfunctions  of 
the operators $ {\hat Q}_{k}$ which we will determine now.
One realizes that the operators ${\hat Q}_{0} $ and 
${\hat Q}_{k} $ only differ by a constant 
term
\begin{displaymath}
{\hat Q}_{k}   = {\hat Q}_{0} + \mu_{k}-\mu_{0}. 
\end{displaymath}
They trivially commute and have
a common set of eigenfunctions $ \psi_{m} $. The eigenvalues
$ \Lambda^{k}_{m} $ of the operators $  {\hat Q}_{k}  $
are related by 
\begin{displaymath}
 \Lambda^{k}_{m}   = \Lambda^{0}_{m} + \mu_{k}-\mu_{0}.
\end{displaymath}
It is, therefore, sufficient to concentrate on the eigenvalue problem
\begin{displaymath}
 {\hat Q}_{0}  \psi_{m} =  \Lambda^{0}_{m}  \psi_{m},
\end{displaymath}
which was studied by Larkin and Ovchinnikov \cite{Lark84} in the
context of single-particle tunneling with dissipation.
They obtained the spectrum
\begin{eqnarray}
&&\Lambda^{0}_{1}=-2 +2 \alpha_c, \nonumber \\
&&\Lambda^{0}_{0}=-2 \alpha_c,\nonumber \\
&&\Lambda^{0}_{-1}=0,\nonumber \\
&&\Lambda^{0}_{m} = 2[1+(|m|-2)\theta/\theta_{0}],  \quad |m| \ge 2,\nonumber
\end{eqnarray} 
where $ \alpha_{c} = 1/2 + \sqrt{5/4 -\theta^{2}/\theta_{0}^{2}}$ and 
showed that the eigenfunctions
\begin{displaymath}
\psi_{m} = \sum_{n=-\infty}^{\infty} C_{m,n} e^{in\tau}   
\end{displaymath}
have Fourier coefficients of the form
\begin{displaymath}
C_{m,n} = \left\{
\begin{array}{ll}
B_{m} ({\tilde C}_{m,n} + d_{m,n}), & n \ge 0, \\
\\
 \pm B_{m}  ({\tilde C}_{m,n} + d_{m,n}), & \pm m>0, \quad n < 0,
\end{array}
\right.
\end{displaymath}
with $ d_{m,n} = 0 $ for $ |m|<2$ and $ |n| +2 > |m| \ge 2$. 
Note that the $ \psi_m $ are even (odd) for positive (negative) $ m $
and the $ B_{m} $ are chosen such that the eigenfunctions are normalized,
\begin{displaymath}
    \langle \psi_{m},\psi_{m} \rangle = 
 \int_{- \pi}^{\pi}{d\tau}~ \psi_{m}^{2}(\tau) = 1. 
\end{displaymath}
For $ m = 0,1 $ they obtained
\begin{equation} \label{Ctilde}
{\tilde C}_{m,n} = 
\left( |n| - \frac{\theta_{0} }{2 \theta}  \Lambda^{0}_{m} \right)
e^{-b|n|}   
\end{equation}
with $\tanh b= \theta/\theta_{0}$. 
In fact, calculating the remaining coefficients for 
$ m \le -1$ and $ m \ge 2$, we find that
Eq.\ (\ref{Ctilde}) holds for any $ m $ with
\begin{displaymath}
d_{m,n}=
\left\{
\begin{array}{ll}
-{\tilde C}_{m,n},                    &|n|<|m|-2, \\
                                      & n=m+2=0,  \\
\\
\frac{1}{2} \left( \frac{\theta_{0}^{2}}{\theta^{2}} + 1 \right),
                                       & n = m-2 = 0, \\
\\
\frac{1}{4} \left( \frac{\theta_{0}^{2}}{\theta^{2}} - 1 \right)
e^{-b(|m|-4)}, & |n| = |m|-2 > 0.
\end{array}
\right.
\end{displaymath}

With these results, we now represent Eq.\ (\ref{inverseinhomo}) in terms of the
basis $ \psi_{m} $ with 
\begin{displaymath} 
 {\tilde q}_{k}^{(i)}= 
\sum_{m=-\infty}^{\infty} c_{k,m}^{(i)} \psi_{m},
\end{displaymath}
\begin{displaymath}
 F_{k}^{(i)} =  \sum_{m=-\infty}^{\infty}  f_{k,m}^{(i)} \psi_{m},
\end{displaymath}
and obtain a special solution in terms of the coefficients 
\begin{equation} \label{coefckm}
c_{k,m}^{(i)}   =  
f_{k,m}^{(i)} / \Lambda^{k}_{m}, \qquad {\rm if}~  \Lambda^{k}_{m}\not=0.
\end{equation}
If  for some $ k' $ and $m'$ the eigenvalue  $\Lambda^{k'}_{m'} = 0$, the
operator $ {\hat Q}_{k} $ is singular and a unique solution of 
Eq.\ (\ref{inhomo}) cannot be found within the $i$th iteration.
However, after performing all necessary iterations, the solutions
have to be the lowest-lying saddle point of the Euclidean action. 
This constraint enables us to  determine the so far 
arbitrary coefficients $ c_{k',m'}^{(i)} $ by requiring
that the Euclidean action as a function of the coefficients
has to be minimal,
\begin{displaymath} 
S_E \left( \left\{ c_{k',m'}^{(i)} \right\} \right) = {\rm min.} 
\end{displaymath}

\section{Nonuniform instanton solution}
After we have explained in detail how to obtain the
approximate solutions, we are now ready to perform the
calculations explicitly. Let us define the parameter 
$ \alpha = (\mu_{1} - \mu_{0})/2 = 1/\sqrt{J}$.
First, we show that the instanton splitting occurs at
$ \alpha = \alpha_{c}$ and  then we apply the perturbation
scheme to determine the split-instanton solutions.

The negative $ \Lambda^{0}_{0} $ indicates that
the operator ${\hat Q}_{0} $ has an unstable mode, which is
responsible for the imaginary part of the free energy and hence for the 
finite decay rate of the metastable state. 
For $ \alpha > \alpha_{c} $, 
the spectrum of $ {\hat Q}_{1} $ is positive
definite. Since $ \mu_{k}-\mu_{0} > 2 \alpha $, for $ k>1$
$\Lambda^{k}_{m} >0$ and hence all higher modes $ q_{k>0}$ are stable.
For $ \alpha < \alpha_{c}$, the lowest eigenvalue 
$ \Lambda_{0}^{1}$ of  $ {\hat Q}_{1} $ 
becomes negative, indicating that the corresponding 
mode also becomes unstable and that a new saddle point 
with a lower $ S_{E} $ exists. Thus at $ \alpha = \alpha_{c}$  
a split-instanton solution emerges. \cite{Mora94}

To determine the split-instanton solutions
for $ \alpha 
{\mathrel{\raise.4ex\hbox{$<$}\kern-0.8em\lower.7ex\hbox{$\sim$}}} 
\alpha_{c} $, 
we now apply the iterative procedure and solve Eqs.\ (\ref{inhomo}). 
In the first iteration $ F_{m}^{(1)} = 0$. According to Eq.\ (\ref{coefckm}) 
most of the coefficients $ c_{k,m}^{(1)} $ are zero except  $ c_{0,-1}^{(1)} $
and $ c_{1,0}^{(1)}$. 
The coefficient $ c_{0,-1}^{(1)} $ cannot be uniquely determined since  
$ \Lambda^{0}_{-1} = 0 $. However, the corresponding
odd eigenfunction $ \psi_{-1} $ is associated to imaginary time 
translation symmetry and does not contribute to the value of $ S_{E} $.
We have, therefore, the freedom to choose $ c_{0,-1}^{(i)} =0$ within
this and the following iterations.
At $ \alpha = \alpha_{c}$, where $ \Lambda_{0}^{1} =0$,
the operator ${\hat Q}_{1} $ becomes singular. Here the instanton splits
since the coefficient $ \zeta \equiv  c_{1,0}^{(1)} $ of the dangerous
mode $ \psi_{0}$ of ${\hat Q}_{1}  $
can have a finite value that remains to be 
determined by minimizing $ S_E(\zeta) $.
To lowest order, the split-instanton solution at $ \alpha = \alpha_{c}$
is thus given by 
\begin{equation} \label{result1}
{\tilde q}_{1}^{(1)}(\tau) =  \zeta \psi_{0}(\tau), 
\qquad {\tilde q}_{k\not=1}^{(1)}(\tau) = 0.
\end{equation}
In analogy with the Landau theory of
phase transitions we can interprete 
$ S_{E}$ as a thermodynamic potential  and $ \zeta $ as 
an order parameter. A
finite $ \zeta $ indicates the existence of a quantum elastic solution.

Using ${\hat Q}_{1}(\alpha) = {\hat Q}_{1}(\alpha_{c})+
2(\alpha -\alpha_{c})$, recalling that 
${\hat Q}_{1}(\alpha_{c}) \psi_1 =0 $ 
and inserting the perturbative result (\ref{result1})
 into Eq.\ (\ref{action2}),
one obtains the split-instanton action up to terms quadratic in 
the dangerous mode, $ S_{E}(\zeta)=B_{qr} +g(\alpha -\alpha_{c})\zeta^{2}$.
However, in order to be able to minimize the action as a function of $ \zeta $
for $ \alpha <\alpha_{c}$, at least 
the terms quartic in $ \zeta $ have to be calculated,
\begin{equation} \label{SE(zeta)}
 S_{E}(\zeta)=
B_{qr} +g \left[ (\alpha-\alpha_{c})\zeta^{2} + \delta \zeta^{4} \right].
\end{equation}
The case $ \delta \le 0$, which indicates that a first-order transition
has occurred will be discussed later in more detail.
For $ \delta>0$, the minimal value of $ S_{E}$ is given by 
$ |\zeta| = \sqrt{ (\alpha_{c}- \alpha)/2 \delta }$ 
and the extremized action reads
\begin{equation} \label{extremizedaction}
 B_{qe}=S_{E}(\zeta)= B_{qr} -\frac{g(\alpha-\alpha_{c})^{2}}{4 \delta}.
\end{equation}
Note  that $ \zeta  $ 
is small close to $ \alpha_{c}$ and can be regarded as a perturbation 
parameter.

In the first iteration we considered corrections to the single 
instanton solution of the order $ \zeta $. In order to determine $ \delta$,
the split-instanton solution up to orders of $ \zeta^{2}$ has to
be treated. 
Consequently, we have to perform the second iteration of the perturbation 
procedure. The inhomogeneous terms $ F_{k}^{(2)} $ in Eq.\ (\ref{inhomo})
are found by substituting $ {\tilde q}_{k}^{(1)} $ into Eq.\ (\ref{force}).
For $ k \not = 0,2$ the $F_{k}^{(2)}= 0$, hence 
$ {\tilde q}_{k}^{(2)} =    {\tilde q}_{k}^{(1)}$, whereas
for  $ k = 0,2$ we obtain $  F_{0}^{(2)}= 2 \zeta^{2} (\psi_{0})^{2} $ 
and  $F_{2}^{(2)}=\sqrt{2} \zeta^{2} (\psi_{0})^{2} $.
The remaining
task within this iteration is to solve the equations 
\begin{eqnarray}
{\hat Q}_{0} {\tilde q}_{0}^{(2)}  &=& 2 \zeta^{2} (\psi_{0})^{2}, 
 \nonumber \\   
{\hat Q}_{2} {\tilde q}_{2}^{(2)}  &=& \sqrt{2} \zeta^{2} (\psi_{0})^{2}. 
 \nonumber
\end{eqnarray}
Representing $ (\psi_{0})^{2} $ in the basis $ \psi_{m} $, we obtain 
 \begin{displaymath}
 (\psi_{0})^{2} =  \sum_{m=0}^{\infty}  a_{m} \psi_{m}.
\end{displaymath}
Note  that the odd $ \psi_{m} $ with $ m<0$ do not appear in the sum, since
$ (\psi_{0})^{2} $ is an even function. 
The coefficients
 $ a_{m} $ are given by  
\begin{displaymath}
 a_{m} = \langle \psi_{m}, \psi_{0}^{2} \rangle 
       =2 \pi \sum_{l,n} C_{m,n} C_{0,l} C_{0,n+l}.
\end{displaymath}

For our purposes
it will be more than sufficient to consider only the first three
coefficients $ a_{0}, a_{1} $ and $ a_{2}$,
since $ a_{m}/a_{m+1} \ll 1 $ in  the entire quantum regime. 
After tedious but straightforward calculations, we find
\end{multicols}
\setlength{\unitlength}{1.cm}
\begin{center}
\begin{picture}(18,0.0)
\put(0.,1.){\line(1,0){8.6}}
\put(8.6,1.){\line(0,1){0.2}}
\end{picture}
\end{center}
\begin{eqnarray}
 a_{0} & =& \label{a0}
\frac{  
        \left( \frac{13}{8}\alpha_{c}+\frac{17}{4} \right)
                \frac{\theta_{0}}{\theta}               
        -\left( \frac{37}{4}\alpha_{c}+ \frac{19}{2} \right)
                \frac{\theta_{0}^{3}}{\theta^{3}}   
        +\left( \frac{69}{8}\alpha_{c}+\frac{21}{4} \right)
                \frac{\theta_{0}^{5}}{\theta^{5}}   
     }
     {\sqrt{2 \pi}
      \left[
        \left( 2\alpha_{c}+ \frac{3}{2} \right)         
                \frac{\theta_{0}^{3}}{\theta^{3}}                       
        -\left( \alpha_{c}+\frac{3}{2}  \right)         
                \frac{\theta_{0}}{\theta}
      \right]^{3/2}
     },
\\  
 a_{1} & =&  \label{a1}
\frac{  
        \left( -\frac{1}{8}\alpha_{c}+\frac{1}{8} \right)
                \frac{\theta_{0}}{\theta}               
        -\left( \frac{1}{4}\alpha_{c}+ \frac{3}{4} \right)
                \frac{\theta_{0}^{3}}{\theta^{3}}   
        +\left( \frac{3}{8}\alpha_{c}+\frac{5}{8} \right)
                \frac{\theta_{0}^{5}}{\theta^{5}}   
     }
     {\sqrt{2 \pi} 
        \left[
        \left(- 2\alpha_{c} + \frac{7}{2} \right)       
                \frac{\theta_{0}^{3}}{\theta^{3}}                       
        -\left( \alpha_{c}-\frac{5}{2}  \right)         
                \frac{\theta_{0}}{\theta}
      \right]^{1/2}
      \left[
        \left( 2\alpha_{c}+ \frac{3}{2} \right)         
                \frac{\theta_{0}^{3}}{\theta^{3}}                       
        -\left( \alpha_{c}+\frac{3}{2}  \right)         
                \frac{\theta_{0}}{\theta}
      \right] 
     },  
\\  
 a_{2} & =& \label{a2}
\frac{  
        \left(  \frac{5}{4}\alpha_{c}+\frac{13}{8} \right)
                \frac{\theta_{0}}{\theta}               
        -\left( \frac{7}{2}\alpha_{c}+ \frac{13}{4} \right)
                \frac{\theta_{0}^{3}}{\theta^{3}}   
        +\left( \frac{9}{4}\alpha_{c}+\frac{13}{8} \right)
                \frac{\theta_{0}^{5}}{\theta^{5}}   
     }
     { \sqrt{2 \pi}
        \left[
               \frac{1}{4} \frac{\theta_{0}^{4}}{\theta^{4}} 
              -\frac{1}{2} \frac{\theta_{0}^{3}}{\theta^{3}}
              +\frac{1}{2} \frac{\theta_{0}^{2}}{\theta^{2}} 
              -\frac{1}{2} \frac{\theta_{0}}{\theta}
              +\frac{1}{4} 
        \right]^{1/2}
      \left[
        \left( 2\alpha_{c}+ \frac{3}{2} \right)         
                \frac{\theta_{0}^{3}}{\theta^{3}}                       
        -\left( \alpha_{c}+\frac{3}{2}  \right)         
                \frac{\theta_{0}}{\theta} 
        \right]
     },\\
 a_{m} & \approx& 0, \qquad m>2. 
 \end{eqnarray}
\begin{center}
\begin{picture}(18,0.0)
\put(9.3,0){\line(1,0){8.5}}
\put(9.3,0){\line(0,-1){0.2}}
\end{picture}
\end{center}
\begin{multicols}{2}
Note that for $ \theta \to \theta_{0}$ the coefficients
converge to $ a_{0} \to 1/\sqrt{2 \pi} $ and $ a_{k} \to 0 $ for $k>0$.
Substituting  $ f_{0,m} = 2\zeta^{2}  a_{m}$ 
and $ f_{2,m} = \sqrt{2}\zeta^{2} a_{m} $ into Eq.\ (\ref{coefckm}) we find 
excellent approximations of the expansion coefficients $ c_{0,m}^{(2)} $ 
and $ c_{2,m}^{(2)}$. Having determined $ {\tilde q}_{0}^{(2)} $ and
 $ {\tilde q}_{2}^{(2)} $, 
the second iteration of the perturbation procedure is completed.  
   
Evaluating Eq.\ (\ref{action2}) with
$ {\tilde q}_{0}^{(2)},{\tilde q}_{2}^{(2)}   \propto \zeta^{2}$,
we obtain the  coefficient $ \delta $ of the quartic term in 
$ S_E(\zeta), $
\begin{equation} \label{intdelta}
\delta =  - \frac{1}{\zeta^{2}}
\int_{-\pi}^{\pi} d\tau  [\psi_{0}(\tau)]^{2}
\left({\tilde q_{0}}^{(2)}+  \frac{{\tilde q_{2}}^{(2)}}{ \sqrt{2}} \right).
\end{equation}
Performing the integral and using the orthogonality of 
\begin{minipage}{8.5cm}
\begin{figure}[t]
\vspace*{-0.3cm}
\begin{center}
  \epsfig{file=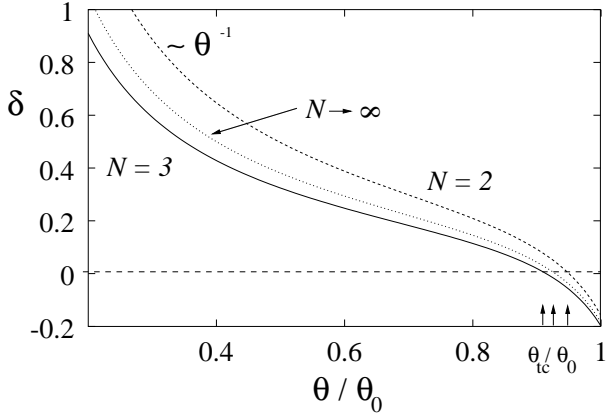,width=8cm}
\vspace*{-0.2cm}
\end{center}
\caption[]{
The fourth-order expansion coefficient $ \delta $ of the Euclidean
action $ S_{E}(\zeta)=B_{qr} 
+g \left[ (\alpha-\alpha_{c})\zeta^{2} + \delta  \zeta^{4} + 
\gamma  \zeta^{6} \right]  $  as 
a function of dimensionless temperature $ \theta /\theta_{0}$ 
for $ N=2,3$ degrees of freedom and the 
continuous limit $ N \to \infty$.
In the low-temperature limit   $ \delta $ is 
positive and diverges as $ \delta(\theta) \sim  1  / \theta $.
At the tricritical temperature $ \theta_{tc}$ it vanishes, 
$ \delta(\theta_{tc}) = 0  $
and becomes negative for $ \theta > \theta_{tc}$. 
The negative $ \delta $ indicates a first-order transition.}
\label{delta}
\vspace{0.5cm}
\end{figure}
\end{minipage}
the $ \psi_{m} $, we find
\begin{equation} \label{sumdelta}
\delta =  -  \sum_{m=0}^{\infty} a_{m}^{2} 
\left( 
\frac{2}{\Lambda_{m}^{0}} + \frac{1}{\Lambda_{m}^{2}}
\right).
\end{equation}
The function $ \delta(\theta/\theta_{0}) $ is shown in Fig.\ \ref{delta} for
$ N=2,3$ and $N \to \infty $. 
At first it may be surprising that 
the curve for $ N \to \infty $ lies
in between the curves for $ N=2 $ and $ N=3$. The reason  is that
for  $ N=2$  the mode $ {\tilde q}_{2} $ does not exist
and $ \delta = -\sum a^{2}_{m}/\Lambda^{0}_{m} $. 
Since  
$ \Lambda^{2}_{m} = \Lambda^{0}_{m} + 8 \alpha_{c} \cos^{2}(\pi/2N) > 0 $
for $N>2$ it is clear that 
$ \delta(\theta/\theta_{0},N=2) > \delta(\theta/\theta_{0},N>2) $.
When evaluating Eq.\ (\ref{sumdelta}) for $ N \ge 3$ one obtains the  relation
$ \delta(\theta/\theta_{0},N) < \delta(\theta/\theta_{0},N+1) $.
In other words,  for  $ N > 3$  the graphs lie in between the ones for
$ N=3$ and $ N \to \infty$.
In the limit $ \theta \to 0$ we 
find  $ \delta \propto\theta_{0}/\theta $ 
when taking into account only the leading terms in
Eqs.\ (\ref{a0})-(\ref{a2}). 
Hence, $ g/\delta$ in Eq.\ (\ref{extremizedaction}) 
converges to a constant value for $ \theta \to 0$. 
With increasing $\theta $ the coefficient $ \delta $ decreases and
vanishes at the characteristic temperature $ \theta_{tc}$, where
Eq.\ (\ref{extremizedaction}) looses its validity.
At $ \theta_{tc }$ the first-order and second-order transition lines merge.
In analogy to the classical theory of phase transitions it is  called 
tricritical temperature. \cite{Ivle87,Mora94} 
Above $ \theta_{tc}$ the parameter $ \delta$
becomes negative, indicating that the transition from rigid quantum
to elastic quantum decay becomes first-order-like.
The values of $ \theta_{tc}$ are given in Table \ref{table}
for $ N=2,3$ and $ N \to \infty$.
  Note  that $ \theta_{tc}$ is smallest for
$ N=3 $ and increases monotonically with $ N$ for $ N>3$.  
Recall that $ \theta_{tc}$ is largest for $N=2$ due 
to the absence of $ {\tilde q}_{2}$ in a system with only two DOF.

We now concentrate on the case where $ \delta < 0 $.
Then in order to find the minimal values of $S_{E}$, 
we have to determine terms of the action $ \propto \zeta^{6}$, 
\begin{equation} \label{action6}
 S_{E}(\zeta)=
B_{qr} +g \left[ (\alpha-\alpha_{c})\zeta^{2} + \delta \zeta^{4} 
+\gamma \zeta^{6}
\right],
\end{equation}
\begin{minipage}[t]{8.5cm}
\vspace*{-0.7cm}
\begin{table}[t]
\begin{tabular}{c|c|c|c}
  $N$                & $2$      & $3$        & $\infty$ \\
\hline 
  $J_{tc}      $    & $ 0.8424 $   & $ 0.7652$   & $ 0.7938 $   \\
\hline 
  $\theta_{tc} $    & $0.8719$   & $0.8000$   & $0.8275$   \\
\hline 
$\delta'( \theta_{tc}/\theta_{0}) $ &   $-2.274$ & $-1.384$ & $-1.675$ \\
\hline 
 $\gamma(\theta_{tc}/\theta_{0})$   &  0.4118 &   0.2105  & 0.2698 \\         
\end{tabular}
\caption{ Numerical values of the tricritical current $ J_{tc} $ 
, tricritical temperature 
$ \theta_{tc} $, the derivative of the coefficient of the forth-order term
$ \delta'(\theta_{tc}/\theta_{0})$ 
and the coefficient $ \gamma(\theta_{tc}/\theta_{0}) $ 
of the sixth-order term for
various numbers of degrees of freedom $N$.
  }
\label{table}
\end{table}
\vspace{0.5cm}
\end{minipage}
which are obtained in the third iteration $(i=3)$
of the perturbation procedure.
For  $ k \not= 1,3$ one has $  {\tilde q}_{k}^{(3)} ={\tilde q}_{k}^{(2)}$.
Inverting the equations 
\begin{eqnarray}
{\hat Q}_{1} {\tilde q}_{1}^{(3)}  &=& 
\left( 4 { \tilde q_{0} }^{(2)}+2 \sqrt{2} {\tilde q_{2}}^{(2)} \right)
\zeta \psi_{0}(\tau), \label{q13} \nonumber \\
{\hat Q}_{3} {\tilde q}_{3}^{(3)}  &=& 
2 \sqrt{2} \zeta \psi_{0}(\tau) {\tilde q}_{2}^{(2)}   \label{q33} \nonumber
\end{eqnarray}
numerically and inserting the values into
 \begin{eqnarray} \label{gamma}
\gamma & =&
\frac{1}{\zeta^{6} }  \int_{-\pi}^{\pi}d\tau 
\left\{
-\sqrt{2}\zeta \psi_{0}(\tau) q_{2}^{(2)}  q_{3}^{(3)}  \right. \nonumber   \\
&&
- \frac{2}{3}  \left( q_{0}^{(2)} \right)^{3} 
- \left( q_{2}^{(2)}\right)^{2}  \left(2 q_{0}^{(2)} 
             -\frac{\sqrt{2} }{3} \delta_{N,3} q_{2}^{(2)} \right) \\
&& \left.
- \left[  q_{1}^{(3)}-\zeta \psi_{0}(\tau)  \right]
 \left[2 q_{0}^{(2)} + \sqrt{2} q_{2}^{(2)} \right]
\zeta  \psi_{0}(\tau) 
\right\},\nonumber
\end{eqnarray}
we calculate the coefficient $ \gamma( \theta/\theta_{0})$.
Minimizing Eq.\ (\ref{action6}) with respect to $ \zeta$ for 
$ \alpha - \alpha_{c} < \delta^{2}/3\gamma$, one obtains, in addition to 
the rigid instanton solution with $ \zeta=0$, an elastic instanton solution 
with
\begin{equation} \label{zeta2}
\zeta^{2}=
 \frac{\delta}{3 \gamma} \left( -1
+ \sqrt{1 - \frac{3 \gamma}{\delta^{2}}(\alpha-\alpha_{c})  }
\right).
\end{equation}
The first-order transition occurs at 
$ \alpha = \alpha_{1} = \alpha_{c} + \delta^{2}/4\gamma $ 
when the nonuniform solution
becomes the global minimum of the action, $ S_{E}(\zeta) = S_{E}(0)=0$.

Using the perturbation scheme, one could, in principle, determine the
split-instanton solution to arbitrary order in $ \zeta $. For our discussion of
the behavior of the decay rate close to the crossover from the single instanton
to the split-instanton regime, the calculation shown above is sufficient.

\section{Results and Discussions}
In this section, we discuss 
the various decay regimes which are presented in the decay diagram 
(Fig.\ \ref{diagram}). Let 
\begin{minipage}[t]{8.5cm}
\begin{figure}[t]
\vspace*{-0.7cm}
\begin{center}
  \epsfig{file=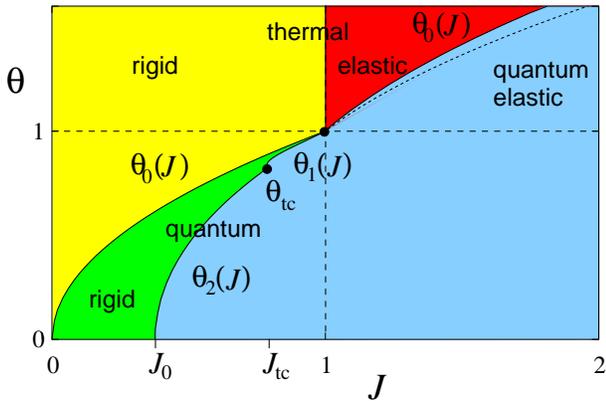,width=8cm}
\end{center}
\vspace*{-0.2cm}
\caption{The decay diagram for $ N=3$ degrees 
 of freedom in terms of the
dimensionless current $ J = \beta^{2}(1-I/NI_{c})/[8 \sin^{4}(\pi/N)]  $ 
and the dimensionless temperature 
$  \theta = \pi \eta \beta T / 2 \hbar E_{J} \sin^{2}(\pi/2N)$.
Criticality $ I = N I_{c} $ corresponds to $ J=0$.
Above $ \theta_{0}(J)$ the system is in the thermal regime, 
and at $ J=1$ a second-order crossover from rigid to elastic thermal decay 
occurs.
At $ \theta_{0}(J)$ a
second-order crossover from thermal to quantum decay takes place.
The quantum regime $ \theta < \theta_{0} $ is again separated into 
rigid and elastic decay. For $ J_{0} < J < J_{tc}$ the crossover from
rigid to elastic decay at $ \theta_{2}(J)$ is of second order. 
For  $ J_{tc} < J < 1$
the crossover indicated by $ \theta_{1}(J)$  is  first-order like. 
Though the diagram is similar for different $ N$, 
the temperatures $ \theta_{tc}$, 
$ \theta_{1}(J)$, and $ \theta_{0}(J>1)$  are 
altered. For comparison, the dashed curve shows  
$ \theta_{0}(J>1)$ for a dc SQUID $(N=2)$.
}
\label{diagram}
\end{figure}
\vspace{0.5cm}
\end{minipage}
us start with the thermal regime $ \theta > \theta_{0}(J) $, where the 
decay occurs via thermal activation. 
For $J < 1$, the coupled DOF behave like a single DOF since the
coupling energy is large compared to the thermal or the barrier energy.
Then the system is in the rigid thermal regime. 
Increasing $J $, 
one enters the elastic thermal decay regime \cite{Droe99}
passing the second-order crossover line at $ {J} =1$.
On the other hand, starting in the thermal rigid  phase and
reducing $ \theta $, quantum fluctuations become important and
at $ \theta_{0}(J)$ a second-order crossover from thermal to quantum decay
takes place. \cite{Lark83}
Two characteristic currents, 
$ J_{0}=0.3820$ and $ J_{tc}$ as given in Table \ref{table}
become important in the quantum regime.
Below $ J_{0}$, the system preferably  decays via the single instanton or  
rigid quantum saddle-point solution. 
For $ J > J_{0} $  a transition from the rigid  
to the elastic quantum decay regime becomes possible.
For  $ J_{0} < J < J_{tc}$, the 
crossover is of second order and is caused by a saddle-point bifurcation
of the Euclidean action occurring at $ \alpha = \alpha_{c}$. 
The dimensionless crossover temperature is then given by
\begin{equation}
\theta_{2} = \left( {J} + \sqrt{{J}} -1 \right)^{1/2}. 
\end{equation}
For $ J_{tc}< J <1 $, 
the crossover is of first
order.
 The transition occurs at $ \alpha = \alpha_{1}$. 
Near $ \theta_{tc}$, we approximate
$ \delta \approx  \delta'(\theta_{tc}/\theta_{0})
[(\theta -\theta_{tc})/\theta_{0}]$ and find that in this limit the crossover 
line is given by
\begin{equation}
\theta_{1}   = \theta_{tc} + \frac{2 \gamma^{1/2}\theta_{0} }
                                 {\delta'(\theta_{tc}/\theta_{0})} 
                            (\alpha-\alpha_{c})^{1/2}. 
\end{equation}
The numerical values for $ \gamma,~ \theta_{tc}$, and $
\delta'(\theta_{tc}/\theta_{0})$ are given in Table \ref{table}.
Note that since $ \partial_{\alpha} \theta_{1} $ diverges as 
$ \alpha \to \alpha_{c}$, the slope of $ \theta_{1}({J})$ is infinite
at the tricritical point. 
For $ J>1$ the transition 
from the thermal to the quantum elastic region is again of second order.
The crossover temperature is then given 
by $ \theta_{0}(J>1)$ [see Eq.\ (\ref{T_0:elastic})].
The decay diagram of an  overdamped JJ array with $ N=3 $ junctions 
presented here is similar to that of the dc SQUID with $N=2$.
Qualitatively, the diagrams exhibit the same features for all $N$.
The reason is that for $ J<1$, the 
transition lines $ \theta_{0}(J)$ and $ \theta_{2}(J)$ are determined
by the long-wavelength modes $ q_{0}^{(0)}$ 
and $ \tilde{q}_{1}$, 
respectively, and hence are independent of $ N$. 
However, there is a difference between the diagrams on a 
quantitative level, since $ \theta_{tc}$,
$ \theta_{1}(J)$ and $ \theta_{0}(J>1)$ are parametrized by $N$. For
example, compared to the dc SQUID, the first-order transition region
is enlarged for $ N>2$ and is the largest for $N=3$.

The remaining task is to discuss the decay rate $ \Gamma \sim \exp(-B/\hbar)$ 
in the four regimes. To exponential accuracy, $ \Gamma $ is determined by 
the extremal of the action $ B $, which is given by the Euclidean action 
$S_{E} $ evaluated at the 
relevant saddle-point configuration $\mbox {\boldmath$\varphi$}_{s}$,
 $ B({J},\theta) = S_{E}[\mbox {\boldmath$\varphi$}_{s}]$.
The behavior of the rate
in the thermal regime was discussed in Ref.\ [9]. 
Since the thermal saddle points $ \mbox {\boldmath$\varphi$}_{ts} $
are independent of imaginary 
time, $ B = \hbar V(\mbox {\boldmath$\varphi$}_{s})/T$ and the rate reduces to 
the classical Arrhenius form, 
$  \Gamma \sim \exp(-V(\mbox {\boldmath$\varphi$}_{s})/T)$. 
In the thermal rigid regime $B$ is given by
\begin{eqnarray}  \label{Btr}
B_{tr} =\frac{2 B_{0} }{3} \frac{{\sqrt {J}}}{\theta},
&&\qquad {J} < 1.
 \end{eqnarray}
Realizing that in the thermal regime 
$ \delta = -(2/\lambda_{0}^{rs}+1/\lambda_{2}^{rs})/(2\pi)$ 
and recalling that $ \alpha=1/\sqrt{J}$, 
one finds, with the help of Eq.\ (\ref{VCO}), the thermal elastic result,
\begin{eqnarray}  \label{Bte}
B_{te} = B_{tr} - g \frac{(\alpha-1)^{2}}{4 \delta}, 
&&\qquad {J} 
{\mathrel{\raise.4ex\hbox{$>$}\kern-0.8em\lower.7ex\hbox{$\sim$}}} 1. 
\end{eqnarray}
In the rigid quantum regime, the action $B_{qr}$ is given by Eq.\ (\ref{Bqr}).
Inserting  Eq.\ (\ref{zeta2}) into Eq.\ (\ref{action6}) we find the extremal 
action in the quantum elastic regime for $ J<1$, 
\begin{eqnarray} \label{Bqe}
B_{qe} &=& B_{qr} - \frac{g}{27 \gamma^{2}} 
\left\{  9 \delta \gamma (\alpha - \alpha_{c}) - 2 \delta^{3} \right.
\nonumber \\
&&\left.  + 2 \left[ \delta^{2} - 3 \gamma( \alpha - \alpha_{c}) \right]^{3/2}
\right\},
\end{eqnarray}
where $ \delta(\theta/\theta_{0})$ is given by Eq.\ (\ref{sumdelta}) and 
$ \gamma(\theta/\theta_{0}) $ was calculated numerically using 
Eq. (\ref{gamma}). The rates for various currents $ {J} $ are 
displayed in Fig. \ref{rate} as a function of temperature.
For $ {J} < {J}_{0}$ the system is in the rigid regime 
for all temperatures $ \theta$ (see Fig.\ \ref{diagram}). 
For $ \theta/\theta_{0} > 1  $ the thermal rigid result (\ref{Btr})
applies. In the rigid quantum 
  \begin{minipage}[t]{8.5cm}
\begin{figure}[t]
\vspace*{-0.7cm}
\begin{center}
  \epsfig{file=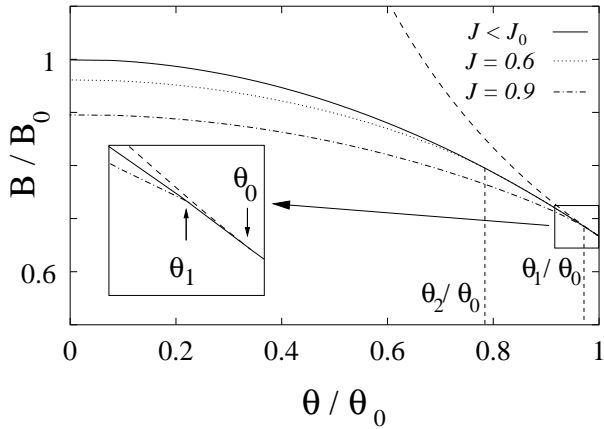,width=8cm}
\end{center}
\vspace*{-0.2cm}
\caption{
The extremal of the action $ B $ 
as a function of temperature $ \theta $ for various normalized currents 
$ {J}$. The dashed line shows the purely thermal behavior
in the rigid regime.
The rigid quantum result is represented by the solid line. 
The dotted and the dashed-dotted lines display  $ B $ of a system with
$ N=3$ degrees of freedom for  ${J}=0.6$ and 
$ {J}=0.9$,
respectively.
For $ {J} =0.6 $ a second-order crossover to the split-instanton
regime occurs at $ \theta_{2}=0.790 \theta_{0} $. 
A first-order crossover takes place for ${J}=0.9 $
at $ \theta_{1}= 0.975 \theta_{0} $. 
The inset shows the cusplike shape of $ B $ close to
$ \theta_{1}$.}
\label{rate}
\vspace*{0.5cm}
\end{figure}
\end{minipage}
regime $ \theta/\theta_{0} < 1  $, in comparison with the purely thermal 
result, the rate is increased due to quantum fluctuations according to  
Eq. (\ref{Bqr}).
In the chosen representation $ B_{qr}$ is independent of system-specific 
parameters. Experimentally measured decay rates of rigid systems  
should thus collapse onto one curve.
For $ {J} > {J}_{0}  $ tunneling of nonuniform 
instantons becomes possible and $ B=B_{qe}$ is reduced further compared to 
$B_{qr} $. In Fig.\ \ref{rate} we displayed $ B $ for a system with $ N=3$ 
degrees of freedom. As an example for the behavior of the rate
close to a second-order crossover to the split-instanton regime we chose 
$ {J} =0.6 < {J}_{tc}$. 
The crossover occurs at $ \theta_{2}=0.790 \theta_{0} $.
For $ {J}=0.9 > {J}_{tc}$, the behavior of the rate is different.
The slope of $ B_{qe}$ changes abruptly at  $ \theta_{1}= 0.975 \theta_{0} $, 
which indicates the occurrence of a first-order crossover.

\section{Conclusions}
In the present work, we studied the decay of metastable states 
in current-driven parallel coupled one-dimensionsal Josephson-junction 
arrays at zero voltage in the overdamped limit. We model this system by 
$ N $ elastically coupled DOF trapped
in the minimum of the single-particle potential
and interacting with a bath of harmonic oscillators. The escape from the
trap can be induced by thermal or quantum fluctuations. Three energy scales
determine the decay behavior of the system; the temperature, the barrier 
height of the trap and the interaction between the particles.
Accordingly, one finds four different regimes for the  decay rate which
we summarized in a decay diagram in Fig.\ \ref{diagram}.
To calculate the decay rate we use the thermodynamic method. 
In the saddle-point approximation, the decay 
is then determined by the most probable configurations leading to an 
escape from the trap which are given by the saddle points of the Euclidean
action. In the thermal regime the saddle-point solutions are independent
of the imaginary time and identical to the saddle points of the potential
energy. \cite{Droe99} If the interaction between the DOF is strong compared
to the barrier energy, rigid configurations dominate the decay. Reducing 
the bias current, the barrier becomes larger and above a critical value the
system preferably decays via an elastic configuration.
On the other hand, starting in the rigid thermal regime and lowering the
temperature, quantum fluctuations become important and the decay 
most probably occurs via the rigid quantum saddle-point or single-instanton 
solution of the Euclidean action. \cite{Lark83}
Inside the quantum region, an elastic regime can again be entered
by increasing the barrier above a critical value. In order to
determine the nonuniform instanton solutions
of the quantum elastic regime, we worked out an iterative perturbation 
procedure. We performed the calculations close to the crossover from rigid
to elastic quantum decay analytically up to second order
and realized the third-order calculation numerically. We were then able to 
give quantitative results for the decay rate including the quantum elastic 
regime. The behavior of the decay rate is similar for SQUID's, DJTL's, and 
long JJ's. In the rigid regime the decay occurs via a saddle point that is 
uniform in space and hence, the qualitative nature of thermal or quantum 
decay is not sensitive to the number of DOF. Further, the 
crossover from rigid to elastic decay is caused by the excitation of 
long-wavelength normal modes of the system, which are equivalent in the three 
physical systems discussed here. 

We want to emphasize that although our conclusions are drawn for overdamped
systems, the reasoning and the procedure also apply for the underdamped case.
Indeed, on a qualitative level, the understanding of the quantum 
rigid-to-elastic crossover in underdamped DJTL's and long JJ's can be 
obtained on the
basis of the theoretical work on SQUID's  (Ref. \cite{Ivle87}).
However, in order to have quantitative results, one has to extend the
theory following the scheme proposed here.

One interesting aspect of the quantum rigid-to-elastic crossover is that
depending on the current, it can be either of first or second order, 
whereas all other crossovers that we discussed are of second order. Even more 
fascinating is the fact
that this crossover is an intrinsic {\em quantum} property and can be
regarded as one further evidence for MQT, if measured. 
Experimental verifications of the predicted enhancement of the decay rate 
due to the elastic properties (see Fig.\ \ref{rate}) are thus highly 
desirable.
An experimental detection of the first-order-like crossover would be 
challenging, but seems to be difficult, 
because the cusplike behavior of the rate at the crossover is not 
very pronounced and occurs in a small current interval. 
In standard experiments, the rate is obtained from the switching current
histogram. The current intervals of the histogram have 
to be much smaller than $ 8 N I_{c} (1-J_{tc}) \sin^{4}(\pi/2N)/\beta^{2} $ 
and the number of events per interval large in order to resolve the 
cusplike feature. Furthermore, it would be convenient to perform the 
measurements on systems with $N = 3$ DOF since in this case the first-order 
region is the largest (see Fig.\ \ref{diagram}). 

In sum, we calculated the decay rate of overdamped  current-biased
one-dimensional Josephson-junction arrays at zero voltage
(including SQUID's, DJTL's, and long JJ's) 
analytically and numerically in several distinct decay regimes.
An experimental observation of
the predicted enhancement of the decay rate in the elastic quantum regime
would give further evidence for macroscopic quantum tunneling in these
systems. 

\section*{Acknowledgments}
We are indebted to   G.~Blatter, O.~S.~Wagner, A.~V.~Ustinov, and A.~Wallraff
for fruitful discussions. Financial support from the DFG-Projekt
No.~Mo815/1-1 and the
Graduiertenkolleg ``Physik nanostrukturierter Festk\"orper,'' University of
Hamburg is gratefully acknowledged.

\bibliographystyle{prsty}

\end{multicols}
\end{document}